\documentclass[12pt,cite,epsf,epsfig]{article}

\usepackage{epsfig}

\begin{document}

\author{Sanjeev Kumar}
\title{Implications of a class of neutrino mass matrices with texture zeros for non-zero $\theta_{13}$}

\date{\textit{Department of Physics and Astrophysics, University of Delhi,\\ Delhi -110007, INDIA.}\\
email: skverma@physics.du.ac.in\\
}

\maketitle

\begin{abstract}

A class of neutrino mass matrices with texture zeros realizable
using the group $Z_3$ within the framework of type (I+II) seesaw 
mechanism naturally admits a non-zero $\theta_{13}$ and 
allows for deviations from maximal mixing. The phenomenology
of this model is reexamined in the light of recent hints for 
non-zero $\theta_{13}$.

\end{abstract}

The observation of neutrino oscillations has provided us 
important information about neutrino masses and mixings. It has 
been known for a long time from the studies of atmospheric 
and solar neutrinos that the mixing angle $\theta_{23}$ 
can be maximal while the  mixing angle $\theta_{12}$ is large but 
well below the maximal value. The smallest mixing angle 
$\theta_{13}$ is allowed to take any value from zero to an upper 
bound given by the accelerator and reactor neutrino experiments. 
However, the global analysis of neutrino data has provided hints 
for a non-zero $\theta_{13}$ at a small statistical significance
not exceeding $2\sigma$ level \cite{fogli08,garcia10,valle11}. 
The first direct observational hint for a non-zero $\theta_{13}$ 
has come from the Tokai-to-Kamioka (T2K) experiment
which has rejected $\theta_{13}=0$ at $2.5\sigma$  \cite{t2k} . 
Shortly after T2K results, the Main Injector Neutrino Oscillation 
Search (MINOS) experiment has disfavored $\theta_{13}=0$ at 
$1.5\sigma$  \cite{minos}. A recent global analysis taking into 
account the recent T2K and MINOS results provides an
evidence for  $\theta_{13}>0$ at a confidence level  
greater than $3\sigma$ \cite{fogli11}. The observation of a
non zero $\theta_{13}$  may further lead to the measurement of 
CP violation in the lepton sector enriching our understanding of 
fermion mass generation and CP violation.

Many successful phenomenological patterns for the neutrino 
mass matrix have been postulated which predict $\theta_{13}=0$ 
as a natural prediction without any fine tuning of parameters. 
These ansatze or symmetries for the 
neutrino mass matrix force certain relations on the elements of
neutrino mass matrix leading to increase in predictability and 
testability. Some popular examples of such ansatze are 
$\mu$-$\tau$ symmetry 
\cite{fukuyama23,mohapatra23,ma23,lam23}, 
TBM mixing \cite{tbm} 
and scaling \cite{scaling}. All these phenomenological 
patterns follow from some discrete flavor symmetries within the 
context of see-saw mechanism. These mass models can explain 
small non-zero values of $\theta_{13}$ as corrections or 
deviations from the exact symmetry. However, it may not be 
possible to accommodate sufficiently large value of $\theta_{13}$ 
in this way. If the recent hints for a large non-zero $\theta_{13}$ 
are confirmed in future, many such ansatze of neutrino mass 
matrix will be overruled. Even if a model survives this 
experimental test, 
one may need to overstretch the model parameters in order
to get a sufficiently large $\theta_{13}$.

Another important phenomenological pattern is neutrino mass 
matrix with texture zeros \cite{zeros} which is also realizable 
from discrete abelian flavor symmetries within the context of 
see-saw mechanism. There are seven patterns of texture zeros,
further divided into three classes (A, B and C), which are 
consistent  with the present experimental data. The phenomenology 
of these classes of neutrino mass matrices has already been 
studied in details \cite{classA, zeros07} 
(and references cited therein).
The main emphasis of the above studies was on finding the 
allowed parameter space and correlations among mixing angles 
and CP violating phases by restricting $\theta_{13}$ less than 6 degrees 
as required by the CHOOZ bound at $1\sigma$ 
\cite{classA, zeros07}. It was found that the classes B and C 
can give a large $\theta_{13}$ only for nearly maximal CP 
violation. This conclusion should not change even after the T2K 
results which point out to larger values of $\theta_{13}$. 
However, the restriction of $\theta_{13}<6$ degrees selects only a 
small portion of the full parameter space and provides only a 
partial understanding of the phenomenology of class A. The aim of 
present communication is to explore the phenomenology of 
class A for non-zero $\theta_{13}$ in the light of recent T2K and
MINOS results.

It is possible to have the symmetry realization of this class of two 
texture zero mass matrices in the context of type (I+II) seesaw  
using the family symmetry $Z_3$ without assuming the 
dominance of either type of contributions keeping only the 
Standard Model scalar doublet which transforms trivially under 
the family symmetry thus effectively suppressing the undesired 
flavor changing neutral currents. An additional advantage of a 
single scalar doublet is the stability of zero textures in the 
neutrino mass matrix under renormalization group evolution. 
These texture zeros are enforced by extending the SM with
one scalar SU(2) triplet. A complex scalar singlet which 
transforms nontrivially under the family symmetry is also 
introduced to get correct type I seesaw 
contribution \cite{z3}.

The neutrino mass matrix $M$ in the flavor basis where
the charged lepton mass matrix is diagonal can be parametrized
in terms of neutrino masses ($m_1$, $m_2$ and $m_3$), mixing 
angles ($\theta_{12}$, $\theta_{23}$ and $\theta_{13}$) and CP 
violating phases ($\alpha$, $\beta$ and $\delta$) as 
\begin{equation}
M=VM_{\nu}^{diag}V^{T}
\end{equation}
where $M_{\nu}^{diag}=Diag \{m_1,m_2,m_3\}$. The mixing 
matrix $V$ is given as
\begin{equation}
V=\left(
\begin{array}{ccc}
c_{12}c_{13} & s_{12}c_{13} & s_{13}e^{-i\delta} \\
-s_{12}c_{23}-c_{12}s_{23}s_{13}e^{i\delta} &
c_{12}c_{23}-s_{12}s_{23}s_{13}e^{i\delta} & s_{23}c_{13} \\
s_{12}s_{23}-c_{12}c_{23}s_{13}e^{i\delta} &
-c_{12}s_{23}-s_{12}c_{23}s_{13}e^{i\delta} & c_{23}c_{13}
\end{array}
\right)\left(
\begin{array}{ccc}
1 & 0 & 0 \\ 0 & e^{i\alpha} & 0 \\ 0 & 0 & e^{i(\beta+\delta)}
\end{array}
\right),
\end{equation}
where $s_{ij}=\sin\theta_{ij}$ and $c_{ij}=\cos\theta_{ij}$. 
The class A of neutrino mass matrices with two texture zeros
has two sub-categories: $A_1$ and $A_2$. In class $A_1$,
$M_{ee}=0$ and $M_{e\mu}=0$ and in class $A_2$,
$M_{ee}=0$ and $M_{e\tau}=0$ \cite{zeros}.

The existence of two texture zeros in the neutrino 
mass matrix give four constraints on the
neutrino mixing parameters. These constraints can be rewritten
in the form of two predictions on the mixing angles $\theta_{13}$ 
and $\theta_{23}$  and two predictions on the CP violating
phases $\beta$ and $\delta$ in terms of the two unknown 
parameters $m_1$ and $\alpha$ \cite{classA} using the 
experimental values \cite{fogli11} of $\theta_{12}$  and
two mass squared differences
$\Delta m^2_{12 }$ and $\Delta m^2_{23 }$ 
 which are used to express $m_2$ and $m_3$ in 
terms of $m_1$. 

The predictions for the two mixing angles in class $A_1$ are 
\cite{classA}
\begin{equation}
\tan^2\theta_{13}=\frac{M}{m_3}
\end{equation}
and
\begin{equation}
\tan^2\theta_{23}=\frac{\mu_3^2 \sin^2 2\theta_{12} }{4 M (M+m_3)}
\end{equation}
where 
\begin{equation}
M=\sqrt{m_1^2 c^4_{12} +m_2^2 s^4_{12} + 2 m_1 m_2
c^2_{12} s^2_{12} \cos 2 \alpha }
\end{equation}
and
\begin{equation}
\mu_3=\sqrt{m_1^2  +m_2^2 - 2 m_1 m_2 \cos 2 \alpha }.
\end{equation}
The mixing angle $\theta_{13}$ is same in class $A_1$ and
class $A_2$. However, the value of $\theta_{23}$ in class
$A_2$ is given by \cite{classA}
\begin{equation}
\tan^2\theta_{23}=\frac{4 M (M+m_3)}{\mu_3^2 \sin^2 2\theta_{12} }.
\end{equation}
The element $m_{ee}$ can only be zero for normal ordering
of neutrino mass spectrum \cite{mee=0}. Therefor, the lightest
neutrino mass is $m_{1}$ which is considered as a free parameter 
whereas $m_{2}$ and $m_{3}$ are expressed in terms of $m_{1}$, 
$\Delta m^{2}_{12}$ and $\Delta m^{2}_{23}$.

The predicted values of $\theta_{23}$ and $\theta_{13}$
span a bounded region on the $(\theta_{23},\theta_{13})$
parameter space depicted in Fig. 1. The dashed (red) line
corresponding to $\alpha=0$ and the dotted (blue) line
corresponding to $\alpha=90$ degrees form the lower boundaries 
of the region while the line $m_1=0.1$ eV forms the upper
boundary of the region. This bounded region is in a way
natural prediction of this class of texture zeros as no fine tuning
of the parameters have been done till now. The experimentally
allowed values of $\theta_{23}$ and $\theta_{13}$ \cite{fogli11}
at $3 \sigma$ C.L. are shown as the solid contours. The predictions
for classes $A_1$ and $A_2$ are mirror images of one another
and are related through the transformation 
$\theta_{23} \rightarrow \pi/2- \theta_{23}$. 
It can be seen from Fig. 1 that $\theta_{13}$ can be
arbitrarily large irrespective of $\alpha$ 
in class $A_1$ ($A_2$) if $\theta_{23}$
is below (above) its maximal value. The only constraint which bounds 
$\theta_{13}$ from above in this region is an upper bound on
the overall neutrino mass scale which has been arbitrarily fixed 
here at $m_1=0.1$ eV for the sake of
illustration. However, the dotted line corresponding to 
$\alpha=90$ degrees forms both lower and upper bounds
on $\theta_{13}$ in the region where $\theta_{23}$ is above
(below) maximality in class $A_1$ ($A_2$).
Moreover, the vanishingly small values of $\theta_{13}$ 
are unnatural in the sense that for these values the deviations
of $\theta_{23}$ from maximality become too large. 
A non-zero value of $\theta_{13}$ consistent with the latest 
neutrino data \cite{fogli11} corresponds to nearly maximal 
$\theta_{23}$ and, therefore, is a natural 
prediction of this class of texture zero mass models.  
The above conclusions become more apparent in Figs. 2 and 3
where the contours of constant $\theta_{13}$ and $\theta_{23}$ 
have been plotted on the $(\alpha, m_1)$ parameter space. 
It is implicit from these figures that the
innermost region where deviations of $\theta_{23}$ from maximal
mixing becomes too large corresponds to $\theta_{13}<6$ 
degrees and, therefore, is disfavored. 

The above analysis has been done for the central values of
$\theta_{12}$, $\Delta m^2_{12}$ and $\Delta m^2_{13}$.
If the errors in these parameters are taken into account, the 
small values of $\theta_{13}$ become marginally allowed 
at $3\sigma$. This can be seen by calculating $\theta_{13}$
and $\theta_{23}$ as functions of $m_1$ for $\alpha=90$ 
degrees by randomly sampling $\theta_{12}$, $\Delta m^2_{12}$ 
and $\Delta m^2_{13}$ \cite{fogli11}. 
The resulting plots have been depicted in Figs. 4 and 5.
The central (red) regions show predictions for the mixing angles 
at $1\sigma$ while the outer (green) regions give the $3\sigma$
predictions. The experimental values \cite{fogli11} at $3\sigma$
have also been depicted for comparison as the horizontal lines.
It is evident from from Fig. 4 that $\theta_{13}$ becomes zero
for $\alpha=90$ degrees at $3\sigma$.
Moreover,  $\theta_{13}$ is non-zero at $1\sigma$ 
and a considerable overlap exits between the model 
predictions and the experimental values for $\alpha=90$ degrees. 

Figs. 3 and 5 also depict the complementarity between
classes  $A_1$ and $A_2$ in their predictions for $\theta_{23}$ in the sense that the predictions of one class are related to the
predictions of the other class by the transformation 
$\theta_{23}\rightarrow \pi/2 - \theta_{23}$.

It has been shown in one of the earlier analyses \cite{classA} that 
predictions of classes $A_1$ and $A_2$ for $\theta_{23}$ 
differ at $3\sigma$  if $\theta_{13}<5$ degrees. In such a
case, $\theta_{23}$ will be above maximal in class 
$A_1$ and below maximal in class $A_2$ \cite{classA}. Since
the new lower bound on $\theta_{13}$ at $3\sigma$ 
is about 4 degrees \cite{fogli11}, such
a distinction between classes $A_1$ and $A_2$ cannot be made
using the latest experimental results. The earlier experimental
data allowed for this distinction between classes $A_1$ and
$A_2$ as there was no lower bound on $\theta_{13}$.

In conclusion, it has been shown that $\theta_{13}=0$ is 
disfavored for neutrino mass matrices of class A since 
this requires $\alpha$ to be fine tuned near $90$ degrees 
which leads to large deviations 
from the maximal value of $\theta_{23}$. If $\theta_{23}$ is
below (above) maximal in class $A_1$ ($A_2$), $\theta_{13}$ can 
be arbitrarily large as it is bounded from above only by the 
upper bound on absolute neutrino mass scale. Therefore,
large values of $\theta_{13}$ can arise naturally in this class
of neutrino mass models. Moreover, it is no longer 
possible to restrict $\theta_{23}$ to above maximality in class
$A_1$ and below maximality in class $A_2$ which was the case 
when there were no lower bounds on $\theta_{13}$.

\newpage

\begin{figure}[tb]
\begin{center}
\rotatebox{0}{\epsfig{file=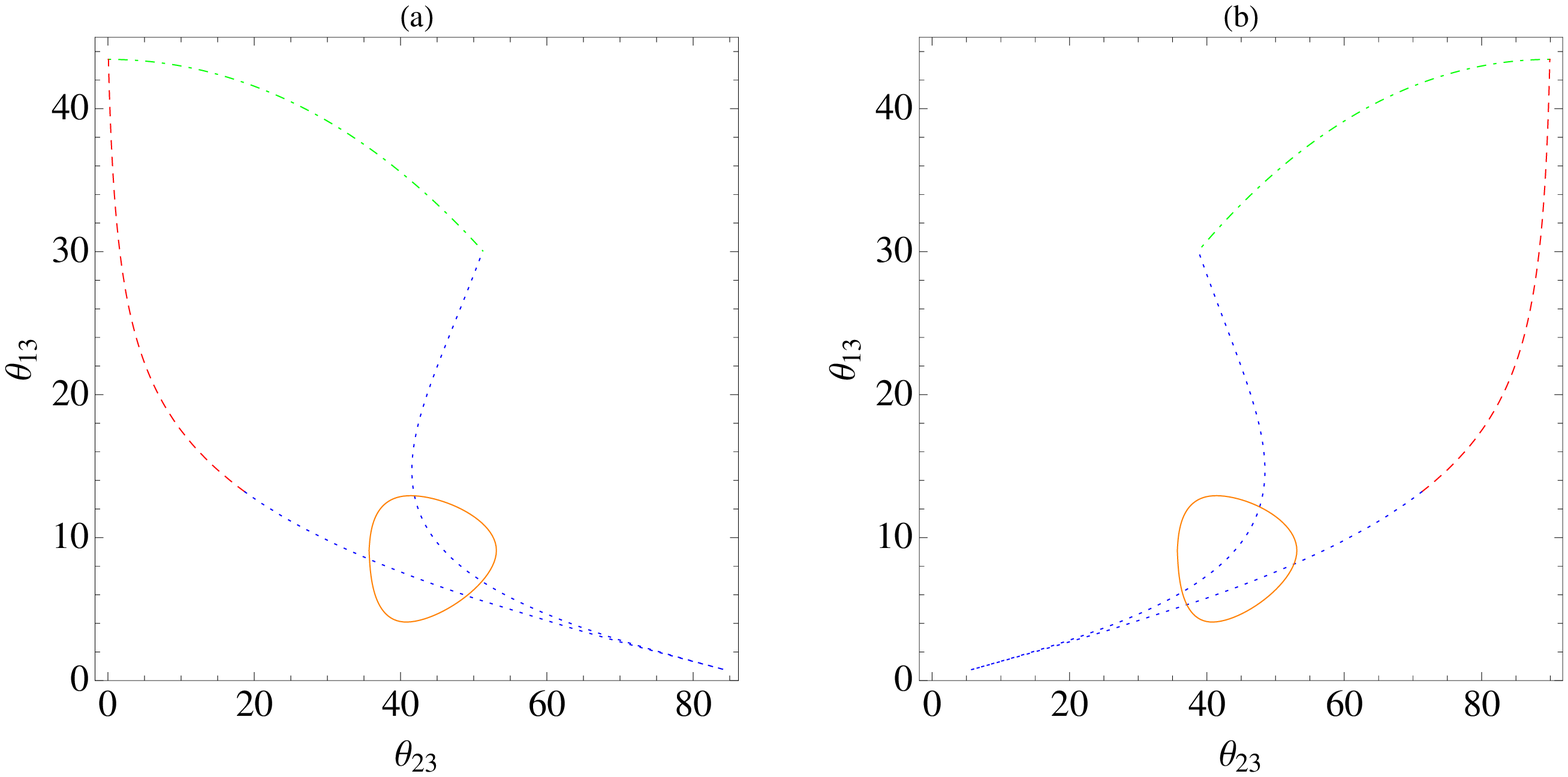, width=16.0cm, height=8.0cm}}

\end{center}

\caption{The regions spanned by predicted values of 
$\theta_{23}$ and $\theta_{13}$ for classes (a) $A_1$ and
(b) $A_2$. The dashed (red) line
corresponds to $\alpha=0$, the dotted (blue) line
corresponds to $\alpha=90$ degrees and the upper
dot-dashed (green) line corresponds to $m_1=0.1$ eV.
The solid (orange) contour shows the $3\sigma$ region
allowed by present experimental data.
}
\end{figure}

\begin{figure}[tb]
\begin{center}
\rotatebox{0}{\epsfig{file=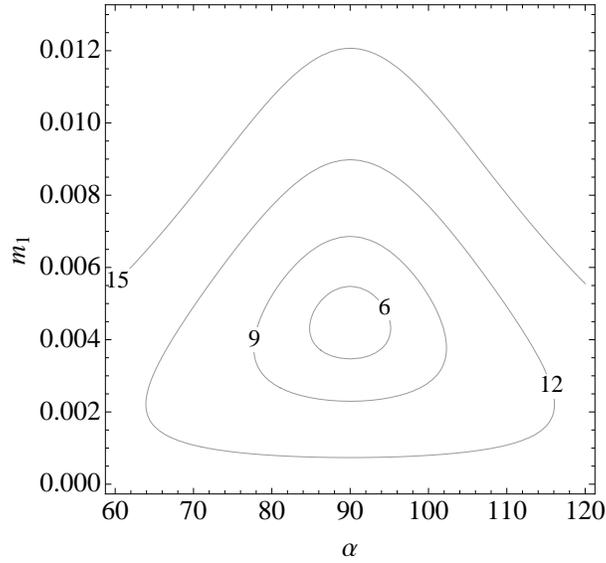, width=8.0cm, height=8.0cm}}

\end{center}
\caption{The contours for constant $\theta_{13}$ on ($\alpha,m_1$) plane.}
\end{figure}

\begin{figure}[tb]
\begin{center}
\rotatebox{0}{\epsfig{file=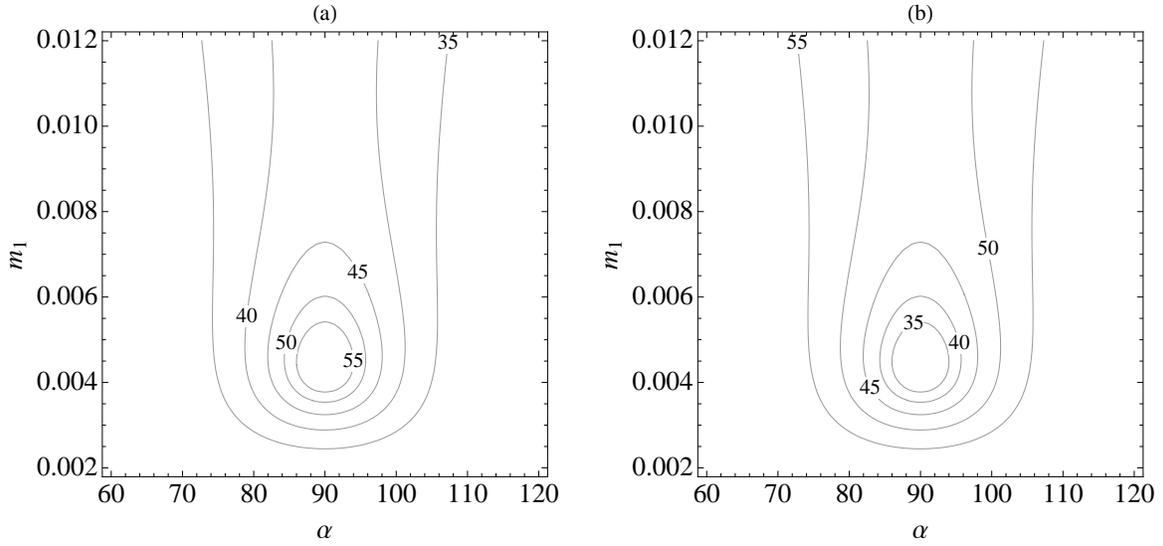, width=16.0cm, height=8.0cm}}

\end{center}
\caption{The contours for constant $\theta_{23}$ on ($\alpha,m_1$) plane for (a) class $A_1$ and
(b) class $A_2$.}
\end{figure}

\begin{figure}[tb]
\begin{center}
\rotatebox{0}{\epsfig{file=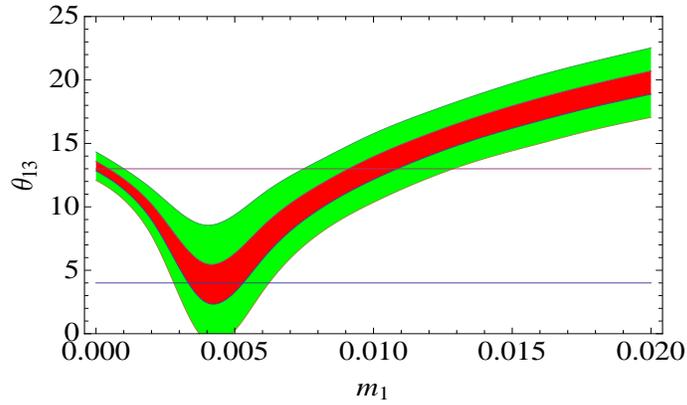, width=9.0cm, height=6.0cm}}

\end{center}
\caption{The variation of $\theta_{13}$ with $m_1$ for
$\alpha=90$ degree. The central (red) region is at $1\sigma$ and the outer (green) region is at $3 \sigma$.}
\end{figure}

\begin{figure}[tb]
\begin{center}
\rotatebox{0}{\epsfig{file=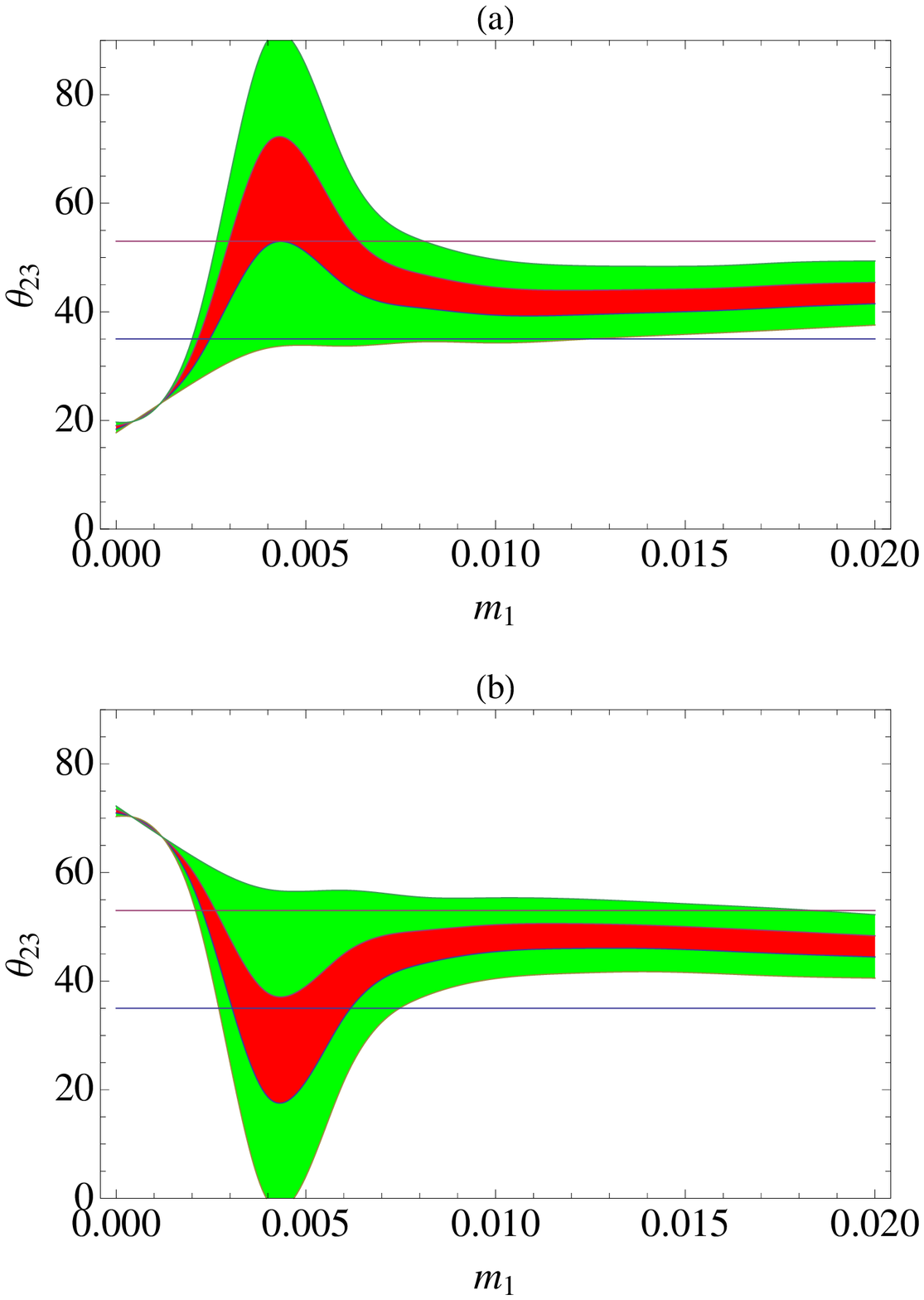, width=14.0cm, height=15.0cm}}

\end{center}
\caption{The variation of $\theta_{23}$ with $m_1$ for
$\alpha=90$ degree for the classes (a) $A_1$ and (b) $A_2$. The central (red) region is at $1\sigma$ and the outer (green) region is at $3 \sigma$.}
\end{figure}


\begin{thebibliography}{99}
\bibitem{fogli08} G. L. Fogli \textit{ et al.}, Phys. Rev. Lett. \textbf{101}, 141801 (2008).

\bibitem{garcia10} M. C. Gonzalez-Garcia, M. Maltoni and
J. Salvodo, JHEP \textbf{04} 056 (2010).

\bibitem{valle11} T. Schwetz, M. Tortola and J. W. F. Valle,
New J. Phys. \textbf{13} 063004 (2011).

\bibitem{t2k} K. Abe \textit{ et al} [T2K Collaboration], 
arXiv:1106.2822 [hep-ex].

\bibitem{minos} L. Whitehead \textit{ et al.} [     MINOS 
Collaboration], Joint Experimental-Theoretical Seminar, (24 June 
2011, Fermilab, USA).

\bibitem{fogli11} G. L. Fogli \textit{et al.}, arXiv:1106.6028 [hep-ph].

\bibitem{fukuyama23} T. Fukuyama and H. Nishiura, hep-ph/9702253.

\bibitem{mohapatra23} R. N. Mohapatra and S. Nussinov, Phys. Rev. D \textbf{60}, 013002 (1999).

\bibitem{ma23} E. Ma and M. Raidal, Phys. Rev. Lett. \textbf{87}, 011802 (2001).

\bibitem{lam23} C. S. Lam, Phys. Lett. B \textbf{507}, 214 (2001).

\bibitem{tbm} P. F. Harrison, D. H. Perkins and W. G. Scott, 
Phys. Lett. B \textbf{530}, 167 (2002).

\bibitem{scaling} Anjan S. Joshipura and Werner Rodejohan, Phys. Lett. B \textbf{678}, 276 (2009).

\bibitem{zeros} Paul H. Frampton, Sheldon L. Glashow and Danny
Marfatia, Phys. Lett. B \textbf{536}, 79 (2002).

\bibitem{classA} S. Dev, Sanjeev Kumar, Surender Verma and Shivani Gupta, Nucl. Phys. B \textbf{784}, 103 (2007).

\bibitem{zeros07} S. Dev, Sanjeev Kumar, Surender Verma and Shivani Gupta, Phys. Rev. D \textbf{76}, 013002 (2007).

\bibitem{z3} S. Dev, Shivani Gupta and Radha Raman Gautam,
Phys. Lett. B. \textbf{701}, 605 (2011).

\bibitem{mee=0} S. Dev and Sanjeev Kumar, Mod. Phys. Lett. A
\textbf{ 22}, 1401 (2007).

\end{thebibliography}
\end{document}